\documentclass[twocolumn,amsmath,amssymb,prb,floatfix]{revtex4}
\usepackage{psfig}
\makeatletter
\usepackage{graphicx}% Include figure files
\begin{document}

\title{On-site Coulomb interaction and the magnetism of (GaMn)N and (GaMn)As.}
\author{
L.M. Sandratskii\thanks{lsandr@mpi-halle.de},$^1$ P. Bruno,$^1$ and  J. Kudrnovsk\'y$^{1,2}$}

\affiliation{$^1$Max-Planck Institut f\"ur Mikrostrukturphysik, D-06120 Halle, Germany\\
$^2$Institute of Physics AS CR, Na Slovance 2, CZ-182 21 Prague, Czech Republic }

\begin{abstract}
We use the local density approximation (LDA) and LDA+U schemes to study
%the influence of Hubbard $U$ on 
the magnetism 
of (GaMn)As and (GaMn)N for a number of Mn concentrations and 
varying number of holes. 
We show that for both systems and both 
calculational schemes the presence of holes is crucial for establishing
ferromagnetism. For both systems, the introduction of $U$ increases
delocalization of the holes and, simultaneously, decreases the 
p-d interaction. Since these two trends exert opposite influences on the 
Mn-Mn exchange interaction the character of the 
variation of the Curie temperature (T$_C$) cannot be
predicted without direct calculation. 
We show that the variation of T$_C$
is different for two systems. For low Mn concentrations we obtain
the tendency to increasing T$_C$ in the case of (GaMn)N whereas an opposite
tendency to decreasing T$_C$ is obtained for (GaMn)As. We reveal the origin
of this difference by inspecting the properties of the densities of states 
and holes for both systems. 
The main body of calculations is performed within a supercell 
approach. The Curie temperatures calculated within the 
coherent potential approximation to atomic disorder are reported
for comparison. Both approaches give similar qualitative
behavior. The results of calculations are related to the experimental data.
%We show that for the same concentration of Mn impurities and the
%same number of holes the Curie temperature can increase for one system and decrease
%for other.  
%For the case of one hole per Mn atom, we obtain for (GaMn)N an 
%increase of T$_C$ for the whole concentration
%range studied: 3\% to 25\%. On the other hand, in (GaMn)As the Curie temperature 
%decreases  for low Mn concentrations and shows a trend to an increase only for very high 
%concentrations. We show that the relation between the Curie temperatures 
%depends on the number of holes.
%exceeding those reachable experimentally. 
%The influence of $U$ on the spatial distribution of the holes, 
%the strength of the p-d interaction
%and the antiferromagnetic superexchange are studied 
%and related to the calculated Curie temperatures. 
%The main body of calculations is performed within a supersell 
%approach. The Curie temperatures calculated within the 
%coherent potential approximation to atomic disorder are reported
%for comparison. Both approaches give similar qualitative
%behavior.
\end{abstract}
\maketitle

\section{Introduction} 

After the discovery of the ferromagnetism of Ga$_{0.947}$Mn$_{0.053}$As
with the Curie temperature as high as 110K \cite{ohno_science}
the diluted magnetic semiconductors (DMS) became the subject of
intensive experimental and theoretical studies being considered as 
promising materials for semiconductor spintronics. Although much new 
understanding of the physics of DMS has been won by the investigation
of (GaMn)As, the Curie temperature of this prototype system could not be raised
above 150K. \cite{ku} The design of technologically useful DMS materials
with the Curie temperature exceeding the room temperature remains
a challenge.

The theoretical prediction by Dietl et al \cite{diohma_science} of the high-T$_C$ 
ferromagnetism in (GaMn)N
played an important role in the formulation of the directions
of the search for new ferromagnetic DMSs. 
A number of experimental
studies indeed detected the ferromagnetism of (GaMn)N samples with the Curie temperature 
higher than the room temperature. \cite{thheov,reelst,soshsa} 
However, the experimental data on the magnetic state of (GaMn)N are strongly scattered 
and range from a paramagnetic
ground state to the ferromagnetic state with a very high Curie temperature of 940 K.

New proof of the complexity of the magnetism of (GaMn)N was given
by the recent magnetic circular dichroism measurements by Ando \cite{ando_gamnn}.
The measurements were performed on a high-T$_C$ sample 
and led Ando to the conclusion that the (GaMn)N phase in this sample
is paramagnetic. The ferromagnetism of the sample was claimed to come from an 
unidentified phase. 
On the other hand, in a recent preprint by Giraud et al \cite{giraud} a high temperature
ferromagnetism was detected in the samples with a low Mn concentration of about 2\%.
The authors rule out the presence of precipitates in the system and argue that
the room temperature ferromagnetism is an intrinsic property of (GaMn)N.
Thus the situation is controversial and calls for further research efforts.
The purpose of the present work is a theoretical comparative study 
of (GaMn)As and (GaMn)N with a focus on the influence of the on-site Coulomb
interaction on the Curie temperature  of both systems. 
The calculations are 
performed on the basis of the density functional theory (DFT) within
both LDA and LDA+U \cite{LDA+U} approach. 
%%?
%A speciall attention is paid to the 
%properties of the hole states mediating the ferromagnetic exchange
%interaction between Mn atoms. 

Numerous LDA calculations of the electronic structure of 
(GaMn)As have been reported 
(see, e.g., reviews Refs. \cite{sato_rev,sanvito_rev} and more recent calculations
Refs. \cite{we_02_03,bokube,bekosa,dederichs}).
%,sanvito_digital}). 
For (GaMn)N the number of such calculations is still small 
(see Refs. \cite{mark01,kulatov,kronik,gamnn_preprint2} and recent preprint 
\cite{gamnn_preprint1}). 
%?
%In general, there is good agreement between different LDA calculations
%concerning the features of the calculated DOS. 
%
The study of the  
properties of both systems which goes beyond the LDA is at the very beginning.
There are a short report on the LDA+U calculation for 
(GaMn)As \cite{gamnas_LDA_U} and a very recent article \cite{gamnn_preprint2} on
LDA+U calculation for both systems. 
We mention also the preprint
on the self-interaction corrected LDA (SIC-LDA) calculation 
\cite{gamnn_preprint1} for these systems. 
%\cite{schulthess-unpublished?} 
%
%In the present paper we report the comparative study of the influence
%of the on-cite Coulomb interaction $_U$ on the ferromagnetism
%in (GaMn)As and (GaMn)N. We perform the LDA and LDA+U calculations
%for both systems and discuss 
%the role of $_U$ in the formation of
%%in the electronic structure, properties of the holes, 
%exchange interactions and Tc. A speciall attention is paid to the 
%properties of the hole states mediating the ferromagnetic exchange
%interaction between Mn atoms. 

The present work is novel in a number of aspects. 
First, we investigate the characteristics 
of the holes and discuss them from the viewpoint of the mediation of the
ferromagnetic exchange interaction between the Mn impurities. 
%Since the holes are commonly considered to be responsible for
%the ferromagnetic ordering their properties provide the basis for the
%discussion of the calculational results on the
%%calculated interatomic 
%exchange interactions and the Curie temperatures. 
Second, we calculate interatomic exchange interactions and the Curie temperatures.
Third, we separate the contributions of the antiferromagnetic superexchange 
%through the filled valence band 
and the ferromagnetic kinetic exchange 
%through the hole states 
to the Curie temperatures. In all studies we focus on the influence of the 
on-site Coulomb interaction.

\section{Calculational scheme}

The calculational scheme is discussed in Refs. \cite{we_02_03,znmnse_03}
to which the reader is referred for more details. The scheme is based on
the DFT calculations for supercells of semiconductor crystals
with one Ga atom replaced by a Mn atom. The size of the supercell 
determines the Mn concentration.    
To calculate the interatomic exchange interactions we use the
frozen-magnon technique and map the results of the total energy of the 
helical magnetic configurations onto a classical Heisenberg Hamiltonian
\begin{equation}
\label{eq:hamiltonian}
H_{eff}=-\sum_{i\ne j} J_{ij} {\bf e}_i\cdot {\bf e}_j
\end{equation}
where $J_{ij}$ is an exchange interaction between two Mn sites $(i,j)$
and ${\bf e}_i$ is the unit vector pointing in the direction 
of the magnetic moment at site $i$.

To estimate the parameters of the Mn-Mn exchange interaction we
perform calculation for the frozen-magnon configurations:
\begin{equation}
\theta_i=const, \:\: \phi_i={\bf q \cdot R}_i
\end{equation}
where $\theta_i$ and $\phi_i$ are the polar and azimuthal angles of vector
${\bf e}_i$, ${\bf R}_i$ is the position of the $i$th Mn atom. 
The directions of the induced moments in the atomic spheres of the atoms of
the semiconductor matrix are kept parallel to the $z$ axis. 

It can be shown that within the Heisenberg model~(\ref{eq:hamiltonian})
the energy of such configurations can be represented in the form
\begin{equation}
\label{eq:e_of_q}
E(\theta,{\bf q})=E_0(\theta)-\sin^2\theta J({\bf q})
\end{equation}
where $E_0$ does not depend on {\bf q} and $J({\bf q})$ is the 
Fourier transform of the parameters of the exchange interaction between 
pairs of Mn atoms:
\begin{equation}
\label{eq:J_q}
J({\bf q})=\sum_{j\ne0} J_{0j}\:\exp(i{\bf q\cdot R_{0j}}).
\end{equation}

Performing back Fourier transformation we obtain the parameters of 
the exchange interaction between Mn atoms:
\begin{equation}
\label{eq:J_0j}
J_{0j}=\frac{1}{N}\sum_{\bf q} \exp(-i{\bf q\cdot R_{0j}})J({\bf q}).
\end{equation}
%The calculation of $E(\theta,{\bf q})$ for different Mn concentrations
%has been performed for uniform meshes in the first BZ. 

The Curie temperature is estimated in the mean-field (MF) approximation
\begin{equation}
\label{eq:Tc_MF}
k_BT_C^{MF}=\frac{2}{3}\sum_{j\ne0}J_{0j}
\end{equation}

We use a rigid band 
approach to calculate the exchange parameters and Curie temperature
for different electron occupations. 
We assume that the electron structure calculated for a DMS with a given
concentration of the 3d impurity is basically preserved in the
presence of defects. The main difference
is in the occupation of the bands and, respectively, 
in the position of the Fermi level.

In the main body of the LDA+U calculations we use $U=0.3$Ry that 
corresponds to the value determined experimentally \cite{okabayashi_jpd} and 
is used in some previous calculations. \cite{gamnas_LDA_U,gamnn_preprint2,
sato_preprint} 
The dependence of the 
Curie temperature on $U$ is illustrated at the end of the paper.
The results of the supercell calculations are compared 
with the results of the calculations within the coherent potential
approximation (CPA). \cite{Turek-book}

\section{Calculational results and discussion}
\subsection{Density of states}

\begin{figure}
\centerline{\includegraphics[width=7.5cm,angle=0]{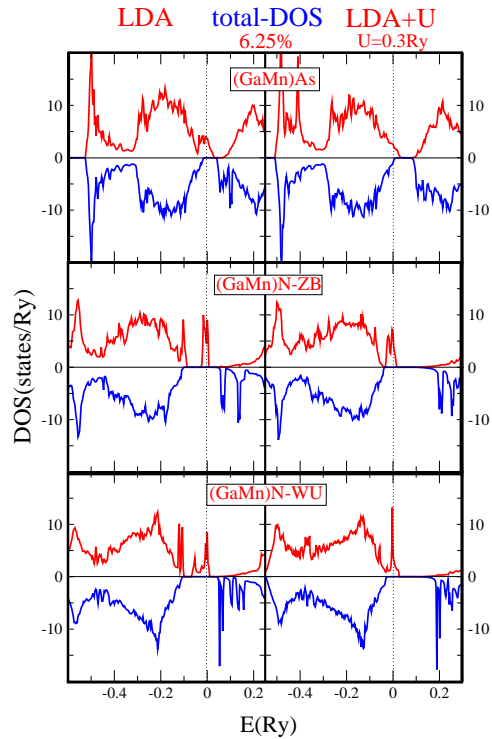}}
%\centerline{\includegraphics[width=7.5cm,angle=0]{fig1_dos_U_tot_Mn.eps}}
\caption{(Color online.) The spin-resolved total 
%and Mn3d 
DOS of (GaMn)As and (GaMn)N for Mn concentration of 6.25\%. 
Calculations are performed within
LDA and LDA+U approaches. For (GaMn)N,
both zinc-blende (ZB) and wurzite (WU) structures are presented.
The spin-up/down DOS is shown above/below the abscissa axis.
The total DOS is given per chemical unit cell of the semiconductor.
%Since one supercell contains 16 semiconductor formula units the 
%Mn3d DOS contributes to the given total DOS with a coefficient $\frac{1}{16}$.
\label{fig_totDOS}}
\end{figure}
\begin{figure}
\centerline{\includegraphics[width=7.5cm,angle=0]{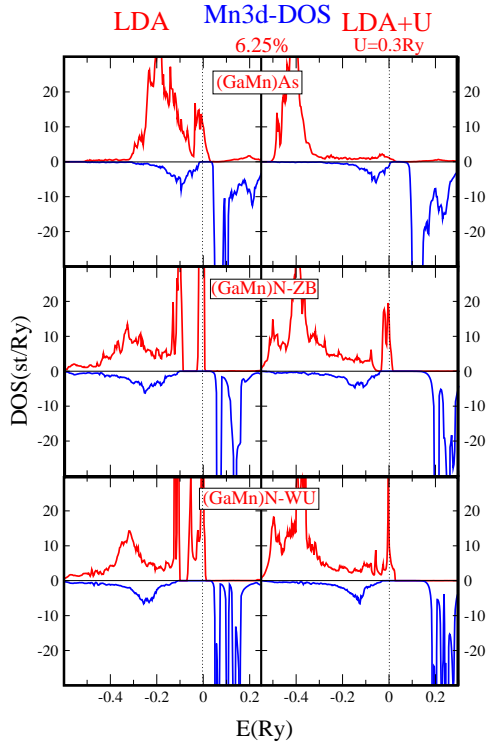}}
\caption{(Color online.) The same as in Fig. \ref{fig_totDOS} but for partial 
Mn3d-DOS.
\label{fig_Mn3dDOS}}
\end{figure}
We begin the discussion 
%of the calculational results 
with the properties of the density of states (DOS). 
In Figs. \ref{fig_totDOS},\ref{fig_Mn3dDOS}
%,\ref{fig_Mn3dDOS} 
we present the DOS for $x=6.25\%$.
The main features of the DOS discussed below are common for 
all Mn concentrations studied. The replacement of
one Ga atom by a Mn atom adds to the energy structure 
five valence bands related to the Mn3d spin-up states. On the other hand, 
there are only four valence electrons added. 
As a result, both (GaMn)As and (GaMn)N have one hole per Mn atom.
(GaMn)N is half-metallic in both LDA and LDA+U calculations with 
only majority-spin states present at the Fermi level. 
(GaMn)As is half-metallic in LDA calculation and possesses small
contribution of the minority-spin states in LDA+U calculation.
In all cases the systems are characterized by high spin-polarization
of the electron states at the Fermi energy. This property is 
in good agreement with a recent measurement of high carrier spin polarization
in (GaMn)As. \cite{brpaxi}  

Despite this similarity, important differences
between (GaMn)As and (GaMn)N are obtained in both LDA and LDA+U calculations.
%(Figs. \ref{fig_totDOS},\ref{fig_Mn3dDOS}). 
In LDA, the spin-up impurity band of (GaMn)As merges with the valence band.
On the other hand, in (GaMn)N the impurity band lies in the semiconducting gap.
This is valid for both wurzite and zinc-blende crystal structures of (GaMn)N.
These features are in agreement with the results of previous calculations.
\cite{sato_rev,sanvito_rev,we_02_03,mark01,kulatov,kronik,gamnn_preprint1,
gamnn_preprint2}
%[
%The width of the impurity band in zinc-blende structure is less than 
%in the wurzite structure. This difference is less pronounced for larger
%Mn concentrations. 
%]
In all LDA calculations the impurity bands have large Mn3d contribution 
(Fig. \ref{fig_Mn3dDOS}).

In the LDA+U calculations, the impurity band of (GaMn)As disappears from
the energy region at the top of the valence band. 
Almost all spin-up Mn3d states lie at about
$-0.4$Ry below the Fermi level. 
%and form a set of lower Hubbard bands.
In contrast, (GaMn)N still possesses impurity band which lies now
closer to the valence band. The Mn3d contribution to the impurity 
band decreases compared with the LDA-DOS but is still large 
(Fig. \ref{fig_totDOS}). 

\subsection{Properties of the holes}

\begin{figure}
%\centerline{\includegraphics[width=5.5cm,angle=0]{fig3_hole_distr_LDA_U_GaMnX.ps}}
\centerline{\includegraphics[width=7.5cm,angle=0]{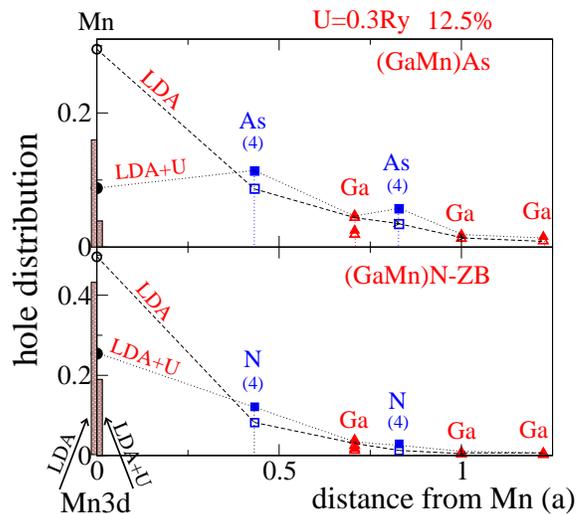}}
\caption{(Color online.) Distribution of the hole in LDA and LDA+U calculations.
Numbers in parentheses give the number of atoms in the coordination sphere.
The hole part is given for one atom.
The distance from Mn is given in units of the lattice parameter. 
%of the ZB structure.  
The narrow rectangulars to the left and right of the ordinate axis present
the Mn3d contribution into the hole for LDA and LDA+U calculations.
%We use this contribution to characterize the strength of the p-d interaction
%for the hole states.
\label{fig_holes}}
\end{figure}
%Since the holes are commonly considered to be
%responsible for the formation of the ferromagnetic
%exchange interactions between the Mn atoms we study next the properties of the holes. 
Now we turn to the discussion of the properties of the holes.
We again present the results for one Mn concentration (Fig. \ref{fig_holes})
and begin the discussion with the spatial distribution of the hole
in the LDA calculations.
%In Fig. we present the properties of the holes states for one Mn concentration. 
%(The calculational data is presented for one Mn concentration of 12.5\%.
%The trends we discuss are valid for other concentrations as well.)
%First we consider the spatial distribution of the hole. We begin with LDA.
In (GaMn)As 29\% of the hole is located on the Mn atom. The first and second 
coordination spheres of As include four atoms each and contain 35\% 
and  14\% of the hole respectively.
%Another ...\% of the hole are located on the second coordination sphere of As.
The rest is distributed between Ga and empty spheres. 

In (GaMn)N the holes are much
stronger localized about the Mn atom. 
%The results are very simular for both ZB and WU crystal structures.
For the ZB structure we find 50\% of the hole in
the Mn sphere. First and second coordination spheres of N contain respectively 
33\% and 5\% of the hole. Although the part of the hole located at the distant
atoms decreases compared with (GaMn)As it is still sizable. This property
allows the mediation of the exchange interaction between the Mn atoms. 
Similar result we obtain for the wurzite (GaMn)N (not shown). 
%(Fig. \ref{fig_holes} ). 

In the LDA+U calculations, the holes become more delocalized (Fig. \ref{fig_holes}). 
In (GaMn)As, the Mn contribution
decreases to 9\%. On the other hand, the contribution on the first and second 
coordination spheres of As
increased to 45\% and 23\%. The contribution of the Ga atoms
changes much less. For (GaMn)N in ZB 
structure the corresponding numbers are 25\%, 48\% and 12\%. 
Again similar behavior is obtained for (GaMn)N in the wurzite structure. 
%(Fig. \ref{fig_holes}).

The increased delocalization of the holes is a factor favorable for
the mediation of the exchange interaction between the Mn atoms and therefore
for a higher Curie temperature. There is however another factor that is
equally important
for the establishing the long-range ferromagnetic order. This factor is the 
strength of the p-d exchange interaction
which is the physical origin of the interaction 
mediated by the holes. 
%between the Mn atoms. 
%This interaction is the physical origin of the effective
%interatomic ferromagnetic interaction between Mn atoms. 
This second factor is less easy to characterize quantitatively. 
%????
A convenient quantity for such a characterization
is the p-d exchange parameter usually denoted as $N\beta$. 
In the mean-field approximation $N\beta$ 
describes the spin-splitting of the semiconductor valence-band states which 
appears as a consequence of the interaction of these states with Mn3d electrons. 
However, our calculations (Fig. \ref{fig_holes}) show that  
the spatial form of the hole states can differ strongly from the unperturbed 
semiconductor states making the mean-field definition of the p-d
interaction poorly founded. 
%Correspondingly, a direct spectroscopical
%determination of $N\beta$ \cite{furdyna} becomes ambiguous 
%since the energy positions of the states involved in the 
%spectroscopic transitions depend on the spatial distribution of the
%hole states and on the admixture of the Mn3d states. 

%Analysis of the experimental data shows that
%for II-VI DMS the value of J_pd is experimentally well established. \cite{}
%For III-V systems the situation is different. 
The experimental estimations of $N\beta$ made for (GaMn)As
on the basis of different experimental techniques
%analysis of the experimental data shows that for (GaMn)As
%the reported values of N\beta 
vary strongly from large values of $|N\beta|=3.3$ eV \cite{matsukura} and 
$N\beta=2.5$ eV \cite{szczytko_SSC} to  much smaller values
$N\beta=-1.2$ eV \cite{okabayashi_jpd} and $|N\beta|=0.6$ eV. \cite{ohakma}
(We use the traditional sign convention: negative $N\beta$
corresponds to antiparallel directions of the spins of the d and p states.) 
A possible reason for this variation is strong p-d hybridization in III-V DMS.
Note that for many II-VI systems $N\beta$ is well defined experimentally 
\cite{furdyna} and 
allows good DFT description within the LDA+U approach. \cite{znmnse_03}
%%the fact that the J_pd parameter is conceptually well defined only in the case of a
%%weak p-d hybridization. \cite{} 
%On the other hand, in the case of a strong p-d interaction
%the hole states differ strongly from the unperturbed semiconductor states and
%contain large contribution of the 3d states (Fig. \ref{}). 
%%In this case the J_pd parameter is not a well defined quantity. 
In the present paper we do not consider the problem of the estimation of 
the $N\beta$ parameter in systems with strong p-d hybridization.
For characterization of the strength of the p-d interaction we use the value of 
the Mn3d contribution to the hole (Fig. \ref{fig_holes}). 
%as the characteristic of the strength of the p-d interaction.

For both systems 
%we obtaine 
the  LDA+U calculations give a strong decrease of the Mn3d contribution 
to the hole (Fig. \ref{fig_holes}). 
%in the case of LDA+U calculation.  
For (GaMn)As, the Mn3d contribution drops 
from  16\% to a small value of 4\%. For 
(GaMn)N in ZB structure we get 43\% and 19\%
and for (GaMn)N in WU structure 41\% and 16\%
(the results for WU structure are not shown in Fig. \ref{fig_holes}). Important that
in (GaMn)N the LDA+U the hole still contains large Mn3d contribution. 
 
The presence of two competing trends in the properties of holes - 
increasing delocalization, on the one hand, and decreasing p-d interaction,
on the other hand - makes direct calculation of T$_C$ necessary
to predict the influence of the Hubbard $U$ on the 
Curie temperature.
%[
%Note that in two opposite limits: of a 3d hole
%localized on the Mn site, on the one hand, and of an undisturbed 
%non spin-polarized semiconductor hole states no ferromagnetic exchange interaction
%is mediated. 
%Here we deal with intermediate cases. 
%] 

\subsection{Curie temperature}

\begin{figure}
%\centerline{\includegraphics[width=5.5cm,angle=0]{Tc_superexch_gamnX_LDA_U.ps}}
%\centerline{\includegraphics[width=7.5cm,angle=0]{fig4_Tc_supercell_CPA.ps}}
\centerline{\includegraphics[width=7.5cm,angle=0]{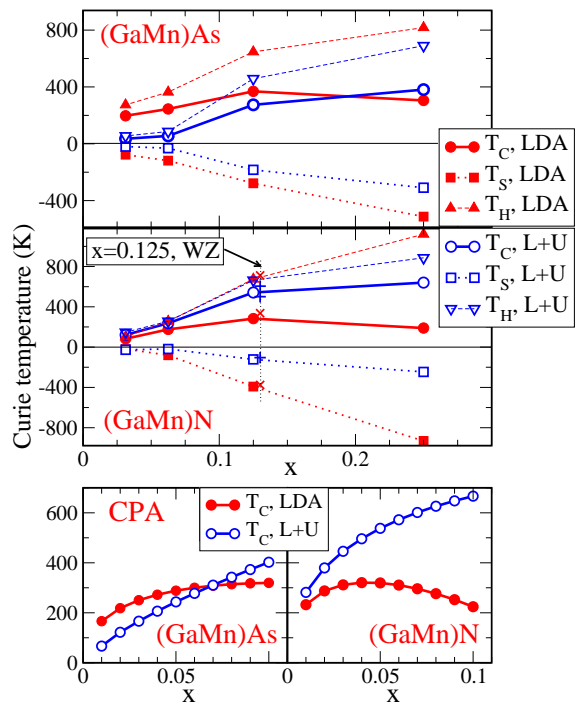}}
\caption{(Color.) T$_C$ as a function of the Mn concentration.
The superexchange contribution T$_{S}$ and hole contribution
T$_H$ are shown (see the text for definitions of these contributions). 
The calculations are performed for ZB structure. For (GaMn)N,
the results for WU structure for $x=12.5\%$ are presented for
comparison. 
The lower part of the figure shows the Curie temperature calculated
within CPA. L+U is abbreviation for LDA+U.
\label{fig_Tc}}
\end{figure}
%revision
In Fig. \ref{fig_Tc}, we show the Curie temperature for a number
of concentrations of Mn impurities. \cite{comment_concentration}

The calculated Curie temperatures (Fig. \ref{fig_Tc}) show again
strong difference 
%of the influence of Hubbard $_U$ for 
between (GaMn)As and (GaMn)N. \cite{comment}
In (GaMn)N, the Curie temperature increases with account for on-site
Coulomb interaction for all Mn concentrations considered. On the other hand,
in (GaMn)As the value of T$_C$ decreases for concentration 3.125\%, 6.25\% and
12.5\%. 
%and increases for 25\%. 

For comparison we present (Fig. \ref{fig_Tc})
the Curie temperature calculated within the 
coherent potential approximation. 
The supercell approach
employed in the present paper and the CPA provide two opposite limits 
in the treatment of the atomic disorder: The supercell calculation deals with an  
ordered atomic pattern. On the other hand, the CPA assumes complete
atomic disorder with the effects of the short range order neglected.
Important that the influence of Hubbard $U$ on the Curie temperature obtained
within CPA is
qualitatively similar to that obtained within the supercell calculations: 
In (GaMn)As, $U$ gives a lower
T$_C$ for small $x$ and a higher T$_C$ for larger $x$. In (GaMn)N,
account for Hubbard $U$ leads to an increase of T$_C$ for the whole
Mn concentration range studied.    
 
\begin{figure}
\centerline{\includegraphics[width=7.5cm,angle=0]{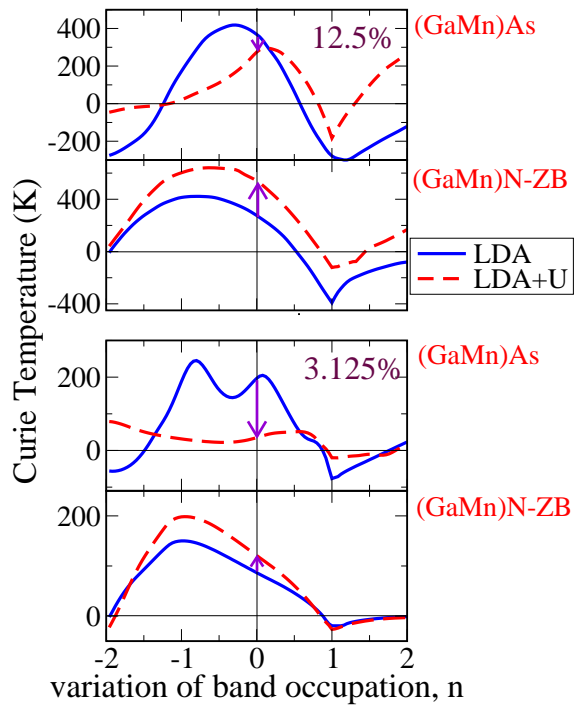}}
%\centerline{\includegraphics[width=7.5cm,angle=0]{Tc_n_gamnx_12_3.eps}}
\caption{(Color online.) The Curie temperature as a function of the band occupation for
Mn concentration of 3.125\% and 12.5\%.
The Curie temperature is calculated according to mean-field formula [Eq. (\ref{eq:Tc_MF})]
and negative values of $T_C$ reflect prevailing antiferromagnetic exchange
interactions.
\label{fig_Tc_n}}
\end{figure}
 
\begin{figure}
\centerline{\includegraphics[width=7.5cm,angle=0]{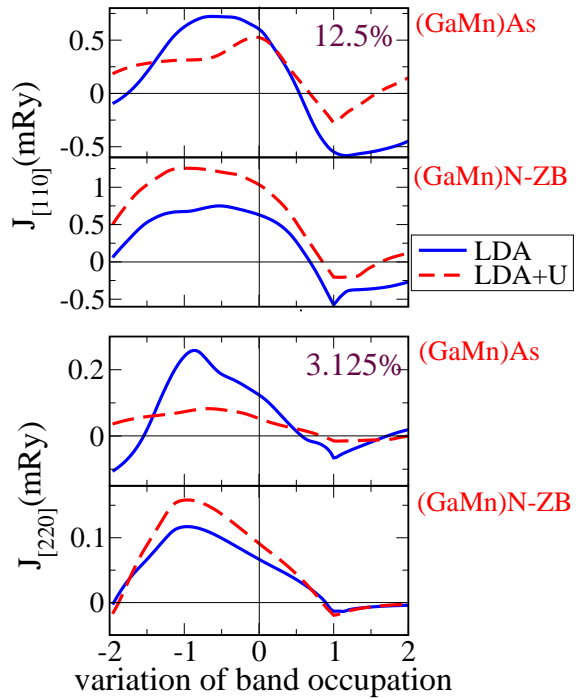}}
%\centerline{\includegraphics[width=7.5cm,angle=0]{J_n_gamnx_12_3.eps}}
\caption{(Color online.) The exchange parameter between Mn atoms separated by the 
vectors $a(1,1,0)$ and $a(2,2,0)$ for, respectively,
$x$=3.125\% and $x$=12.5\%  as a function of the band occupation.
$a$ is the lattice parameter of the zinc-blende structure.
\label{fig_J_n}}
\end{figure} 

To get a deeper insight into the formation of the Curie temperature 
we study the dependence of T$_C$ on the band occupation.  
In Fig. \ref{fig_Tc_n} we present the results for two Mn concentrations
of 3.125\% and 12.5\%. Here, $n=0$ corresponds to the nominal number of carriers 
(one hole per Mn atom).
The kink at $n=1$ marks the point where the valence band is full and the 
conduction band is empty. In  Fig. \ref{fig_J_n}, we show the dependence
of the leading interatomic exchange parameter \cite{we_02_03} as a function of the band 
occupation. Indeed, the behavior of the Curie temperature (Fig.  \ref{fig_Tc_n})
repeats in gross features the behavior of the exchange parameter
(Fig.  \ref{fig_J_n}). In particular, the decrease
of the number of holes below the nominal value of one hole per Mn atom
leads to the decrease and change of the sign of the exchange parameter. 
%revision
Thus our calculations show the crucial role of the valence-band holes 
for mediating ferromagnetism in both (GaMn)As and (GaMn)N. This result
is obtained in both LDA and LDA+U approaches and agrees with the commonly 
accepted picture of the ferromagnetism in DMS used in the model-Hamiltonian
studies (see, e.g., reviews \cite{dietl_rev,kacman_review} and references 
therein).   

It is useful to present the Curie temperature
in the form T$_C$=T$_{S}$+T$_H$. Here T$_{S}$ is the Curie temperature
at $n=1$ and specifies the exchange interaction mediated by the completely filled
valence band. This contribution can be related to the Anderson's
superexchange. \cite{comment1}
T$_H$ is the difference between the Curie temperatures 
at $n=1$ and $n=0$ and specifies the contribution of the hole states.  
%The values of T$_{S}$ and T$_H$ are also presented in Fig. . 

The value of T$_{S}$ is negative for all concentrations
in both LDA and LDA+U approaches
reflecting the property that the exchange interaction through completely filled bands
is antiferromagnetic. 
The hole contribution T$_H$ is always positive and is responsible
for the ferromagnetism of the system. 

The analysis of Fig. \ref{fig_Tc} shows that the contribution of the antiferromagnetic 
superexchange is in general not small and must be taken 
into account in the estimation of the Curie temperature. The absolute value 
of the antiferromagnetic superexchange is always larger in the LDA case. 
The decrease of the superexchange for nonzero $U$ can be explained by an increased
energy distance between the occupied and empty Mn3d states. 

The influence of Hubbard $U$ on the hole 
contribution T$_H$ differs strongly for two systems. 
In (GaMn)N, LDA and LDA+U give very similar results up to x=12.5\%.
For larger $x$ the LDA value of T$_C$ is higher.
This shows that 
for low Mn concentrations the increased delocalization of the holes is compensated by the 
decreased p-d interaction. For larger $x$ the increase of the p-d interaction 
becomes more important
since the distance between the Mn atoms decreases.
For (GaMn)As, the strong decrease of the p-d interaction
under the influence of Hubbard $U$ is not compensated by 
the increased hole delocalization and the value of T$_H$ 
drops strongly for small $x$.  

Comparison of the results of the calculations with experiment shows 
that in the case of (GaMn)As the account for Hubbard $U$ does not make the agreement 
with experiment better. Indeed, the calculated Curie temperature 
becomes too low.
A possible reason for this can be an underestimation of the Mn3d contribution
into the hole states within the LDA+U calculation:
these states disappear from the Fermi-level region almost completely 
(Fig.~\ref{fig_totDOS}).
%decreases 
%too strongly the contribution of the 3d states to the region about the Fermi level. 
The dynamical mean-field theory (DMFT) which treats the on-site correlations 
beyond the LDA+U can improve the
agreement with experiment since the DMFT provides an account for the
quasiparticle states \cite{sakoab} that are neglected within the LDA+U.
%(and, most probably,  overestimated by LDA).
%revision
Before the LDA+DMFT study is performed the decision about more appropriate 
method for the study of the ferromagnetism in (GaMn)As should be made in
favor of the LDA scheme which gives good correlation with experiment. \cite{we_02_03}

In (GaMn)N, the Mn3d states are present at the Fermi level because of the strong
hybridization with the N2p states. 
%Therefore the LDA+U is in this case an
%adequate approach to the account for 
%on-site correlations. 
We obtained an increase of T$_C$
which is in correlation with the prediction by Dietl et al \cite{diohma_science}
and with some of the
experimental results. Our LDA+U results for (GaMn)N correlate in many respects
with the picture proposed recently by Dietl et al \cite{dimaoh} 
in the discussion of the
Zhang-Rice limit for this system. They conclude that the high T$_C$ can appear
in (GaMn)N as a result of strong p-d interaction within the Zhang-Rice polaron
which is efficiently transfered due to the relatively large extent of 
the p-wave functions. \cite{comment_models}

\begin{figure}
\centerline{\includegraphics[width=7.5cm,angle=0]{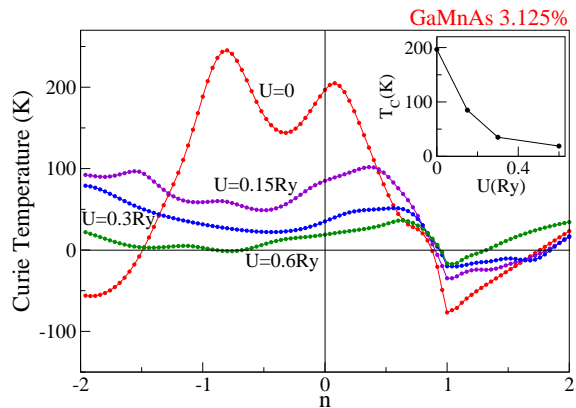}}
\caption{(Color online.) The Curie temperature of (GaMn)As with 3.125\% of Mn as a function of the 
band occupation for a number of $U$ values. In the insert, the $T_C$ value
as a function of $U$ for nominal electron number $n=0$.
\label{fig_varU}}
\end{figure}

Figure \ref{fig_varU} illustrates the dependence of the Curie temperature 
on the value of U for (GaMn)As with 3.125\% of Mn. We obtain a systematic 
trend of decreasing
scale of the variation of $T_C$ with variation of the band occupation.
The insert shows $T_C(U)$ for nominal electron number $n=0$. The
dependence is monotonous though non-linear. 

Within our LDA+U calculation we did not obtain the 
orbital ordering found in the SIC calculation. \cite{gamnn_preprint1}
We, however, do not exclude
that such a state can be stabilized within the LDA+U approach. Our numerical
experiment with the impurity band split "by hand" to simulate the effect 
reported in Ref. \cite{gamnn_preprint1} indeed resulted in a nonferromagnetic 
ground state.

%revision
%Before a better characterization of the (GaMn)N samples will be achieved
%and more coherent experimental information on magnetism of this system is
%available it has a rather restricted usefulness to compare the theoretical
%results with a particular experiment. 
%Further experimental and theoretical efforts are needed to understand the 
%scattering of the experimental
%data for (GaMn)N. One of the possible explanations 
%can be the sensitivity of the system to the screening of the on-site
%Coulomb interaction which can vary from sample to sample or even from 
%Mn atom to Mn atom in the same sample. 

%Another important factor can be the clustering of the Mn atoms.

\section{Conclusions}
We use the local density approximation (LDA) and LDA+U schemes to study
the magnetism of (GaMn)As and (GaMn)N. 
%revision
We show that for both systems and both 
calculational schemes the presence of holes is crucial for establishing
ferromagnetism.
The introduction of $U$ increases
delocalization of the holes and, simultaneously, decreases the 
p-d interaction. Since these two trends exert opposite influences on the 
Mn-Mn exchange interaction the character of the 
variation of the Curie temperature (T$_C$) cannot be
predicted without direct calculation. 
We show that the variation of T$_C$
is very different for two systems. 
For low Mn concentrations we obtain
the tendency to increasing T$_C$ in the case of (GaMn)N whereas an opposite
tendency to decreasing T$_C$ is obtained for (GaMn)As. We reveal the origin
of this difference by inspecting the properties of the densities of states 
and holes for both systems. 
The main body of calculations is performed within a supercell 
approach. The Curie temperatures calculated within the 
coherent potential approximation to atomic disorder are reported
for comparison. Both approaches give similar qualitative
behavior. We make a contact between calculational results
and the strong scattering of the experimental data on T$_C$ of the systems.

Note that the present experimental and theoretical situation does
not allow to give a unique answer on the question which of two schemes, LDA or
LDA+U, provides better description of the diluted magnetic
semiconductors. This answer can be different for different
systems and even samples. Our previous calculations for the II-VI DMS (ZnMn)Se \cite{znmnse_03}
gave strong arguments for the superiority of the LDA+U approach in this case. 
In the present study of (GaMn)As, the LDA+U method with a moderate $U$  
underestimates the strength of the p-d interaction and leads to too 
low Curie temperature
for low Mn concentrations. For (GaMn)N, the experimental situation is too 
unclear to allow for a well-founded decision concerning the superiority of the
theoretical scheme. The comparative study reported here aims to contribute to both 
the increase of the understanding of the systems studied and the formulation
of the theoretical approaches most suitable for the investigation of
concrete DMSs. The efforts must, however, be continued and we hope
that the results of this paper will be stimulating for further studies.

\begin{acknowledgments}
L.S. and P.B. acknowledge the support from Bundesministerium 
f\"ur Bildung und Forschung. J.K. acknowledges the support from 
the Grant Agency of the AS CR
(A1010203) and the RTN project (HPRN-CT-2000-00143). 
\end{acknowledgments}


\begin{thebibliography}{99}
\bibitem{ohno_science}
H. Ohno, Science {\bf 281}, 951 (1998).
\bibitem{ku}
K. C. Ku, S. J. Potashnik, R. F. Wang, S. H. Chun, P. Schiffer, 
N. Samarth, M. J. Seong, A. Mascarenhas, E. Johnston-Halperin, R. C. Myers, A. C. Gossard, 
and D. D. Awschalom, Appl. Phys. Lett. {\bf 82}, 2302 (2003). 
\bibitem{diohma_science}
%T. Dietl et al, 
T. Dietl, H. Ohno, F. Matsukura, J. Cibert, and D. Ferrand, 
Science {\bf 287}, 1019 (2000) 
\bibitem{thheov}
%N. Theodoropoulou et al,  Appl. Phys. Lett. {\bf 78}, 3475 (2001). 
N. Theodoropoulou, A. F. Hebard, M. E. Overberg, C. R. Abernathy, S. J. Pearton, S. N. G. Chu, and R. G. Wilson,  
Appl. Phys. Lett. {\bf 78} 3475 (2001). 
\bibitem{reelst}
%M. L. Reed et al,  Appl. Phys. Lett. {\bf 79}, 3473 (2001).
M. L. Reed, N. A. El-Masry, H. H. Stadelmaier, M. K. Ritums, M. J. Reed, 
C. A. Parker, J. C. Roberts and S. M. Bedair,   Appl. Phys. Lett. {\bf 79}, 3473 (2001).
%
\bibitem{soshsa}
%S. Sonoda et al, J. Cryst. Growth  {\bf 237}, 1358 ((2002).
S. Sonoda, S. Shimizu, T. Sasaki, Y. Yamamoto, and H. Hori,
J. Cryst. Growth  {\bf 237}, 1358 ((2002).
\bibitem{ando_gamnn}
K. Ando, Appl. Phys. Lett. {\bf 82}, 100 (2003).
\bibitem{giraud} 
%R. Giraud et al, cond-mat/0307395
R. Giraud, S. Kuroda, S. Marcet, E. Bellet-Amalric, X. Biquard, 
B. Barbara, D. Fruchart, D. Ferrand, J. Cibert, H. Mariette, cond-mat/0307395
\bibitem{LDA+U}
%V. I. Anisimov et al, 
V.I. Anisimov, J. Zaanen, and O.K. Andersen, Phys. Rev. B {\bf 44}, 943 (1991);
V. I. Anisimov, I. V. Solovyev, M. A. Korotin, M. T. Czyzyk, and G. A. Sawatzky, 
Phys. Rev. B {\bf 48}, 16929 (1993);
V. I. Anisimov, F. Aryasetiawan, and A. I. Lichtenstein, 
J. Phys.: Condens. Matter {\bf 9}, 767 (1997).
%
\bibitem{sato_rev}
K. Sato, and H. Katayama-Yosida, Semicond. Sci. Technol. {\bf 17}, 367 (2002). 
\bibitem{sanvito_rev}
S. Sanvito, G. J. Theurich and N. A. Hill,
J. Superconductivity 15, 85 (2002).
\bibitem{we_02_03} 
L. M. Sandratskii and P. Bruno, Phys. Rev. B {\bf 66}, 134435 (2002);
 Phys. Rev. B {\bf 67}, 214402 (2003).
\bibitem{bokube}
%G. Bouzerar et al, 
G. Bouzerar, J. Kudrnovsky, L. Bergqvist, and P. Bruno,
Phys. Rev. B {\bf 68}, 081203 (2003).  
\bibitem{bekosa} 
%L. Bergqvist et al,
L. Bergqvist, P. A. Korzhavyi, B. Sanyal, S. Mirbt, I. A. Abrikosov, 
L. Nordstr\"om, E. A. Smirnova, P. Mohn, P. Svedlindh, and O. Eriksson,
Phys. Rev. B {\bf 67}, 205201 (2003). 
\bibitem{dederichs}
K. Sato, P. H. Dederics, and H. Katayama-Yoshida,   
Europhys. Lett. {\bf  61} 403 (2003).
%\bibitem{sanvito_digital}
%S. Sanvito, Phys. Rev. B {\bf 68}, 054425 (2003) 
\bibitem{mark01}
M. van Schilfgaarde and O. N. Mryasov, Phys. Rev. B {\bf 63} , 233205 (2001).
\bibitem{kulatov}
%E. Kulatov et al, 
E. Kulatov, H. Nakayama, H. Mariette, H. Ohta, and Yu. A. Uspenskii,
Phys. Rev. B {\bf 66}, 045203 (2002).
\bibitem{kronik}
L. Kronik, M. Jain, and J. R. Chelikowsky
Phys. Rev. B {\bf  66}, 041203 (2002). 
\bibitem{gamnn_preprint2}
B. Sanyal, O. Bengone, and S. Mirbt,  Phys. Rev. B {\bf 68}, 205210 (2003).
\bibitem{gamnn_preprint1}
A. Filippetti, N. A. Spaldin, and S. Sanvito, cond-mat/0302178. 
\bibitem{gamnas_LDA_U}
J. H. Park, S. K. Kwon, and B. I. Min, Physica B {\bf 281-282},
703 (2000).
%
\bibitem{znmnse_03}
L. M. Sandratskii,  Phys. Rev. B {\bf 68}, 224432 (2003).

\bibitem{okabayashi_jpd}
%J. Okabayashi et al, 
J. Okabayashi, A. Kimura, O. Rader, T. Mizokawa, A. Fujimori, T. Hayashi, and M. Tanaka,
Phys. Rev. B {\bf 58}, 4211 (1998).
%

\bibitem{sato_preprint}
K. Sato, P. H. Dederichs, H. Katayama-Yoshida, J. Kudrnovsk\'y (unpunlished).
%
\bibitem{Turek-book}
I.~Turek, V.~Drchal, J.~Kudrnovsk\'{y}, M.~\v{S}ob, and 
P. Weinberger, {\it Electronic Structure of Disordered Alloys, 
Surfaces and Interfaces}, (Kluwer, Boston, 1997).
%
\bibitem{brpaxi}
J. Braden, J. S. Parker, P. Xiong, S. H.  Chun and N. Samarth, 
Phys. Rev. Lett. {\bf 91}, 056602 (2003).

\bibitem{matsukura}
%F. Matsukura et al, 
F. Matsukura, H. Ohno, A. Shen, and Y. Sugawara,
Phys. Rev. B {\bf 57}, 2037 (1998).
\bibitem{szczytko_SSC}
%J. Szczytko et al, 
J. Szczytko, W. Mac, A. Stachow, A. Twardowski, P. Becla, and J. Tworzydlo , 
Solid State Commun. {\bf 99}, 927 (1996).


\bibitem{ohakma}
H. Ohno, N. Akiba, F. Matsukura, A. Shen, K. Ohtani, and Y. Ohno,
Appl. Phys. Lett. {\bf 73}, 363 (1998).
%
\bibitem{furdyna}
J. K. Furdyna, J. Appl. Phys. {\bf 64}, R29 (1988).
\bibitem{comment_concentration}
The calculation is performed for the case of one hole per Mn atom.
Many experiments report the hole number smaller than the 
nominal Mn content. The reason for decreased hole number are
different types of donor defects compensating the Mn acceptors.
On the other hand, the number of the Mn atoms
participating in the ferromagnetic ordering is also found experimentally
to be smaller than the nominal Mn concentration. 
Therefore the ratio of one hole per one active Mn can
still correspond to the factual relation in some samples. The 
scattering of the experimental data discussed in Introduction
reveals differences in nominally
equivalent samples with equal Mn content. To address this problem
we consider the variation of the number of
carriers for a given Mn concentration.

\bibitem{comment}
A. Janotti, S.-H. Wei, and L. Bellaiche [Appl. Phys. Lett. {\bf 82}, 766 (2003)]
report LDA calculation of the total energy of the ferromagnetic and 
antiferromagnetic zinc-blende binary compound MnN and find the 
ground state to be antiferromagnetic. Our calculation for MnN (not shown) 
agrees with this result.
Both the comparison of the total energies of the ferromagnetic and 
antiferromagnetic structures and the frozen-magnon calculation of the
interatomic exchange parameters reveal prevailing character of the
antiferromagnetic exchange interactions in MnN. The situation changes, however,
in the case of (GaMn)N discussed in this paper. Here the leading exchange parameters 
are ferromagnetic.
%

\bibitem{dietl_rev}
T. Dietl, Semicond. Sci. Technol. {\bf 17}, 377 (2002).
\bibitem{kacman_review}
P. Kacman, Semicond. Sci. Technol. {\bf 16}, 25 (2001).

\bibitem{comment1}
Note that the treatment of T$_{S}$ contribution to the Curie temperature as
Anderson's superexchange assumes a rigid band picture where the filling 
of the hole states does not change the electron structure. In many cases 
the rigid band picture is qualitatively valid [see, e.g., 
L. M. Sandratskii and P. Bruno, J. Phys.: Cond. Matt. 
{\bf 15} L585 (2003)]. 

\bibitem{sakoab}
S. Y. Savrasov, G. Kotliar, and E. Abrahams,
Nature {\bf 410}, 793 (2001).

\bibitem{dimaoh}
T. Dietl, F. Matsukura, and H. Ohno, Phys. Rev. B {\bf 66}, 33203 (2002).


\bibitem{comment_models}
At this point it is worth to remind the reader about the strong scattering of the
experimental data discussed in the Introduction. Any theoretical model, also
the models used in the given paper, does not take into account the whole diversity of
the features characteristic for a given sample. The final effect of
the trends revealed in our study in the case of a concrete sample depends on
the combination of defects in the sample. As follows from our study, the hole number
and the efficiency of the screening of the on-site Coulomb interaction are
important characteristics. Another feature which can play an important role
in the magnetism of DMS is clustering of the magnetic impurities. \cite{mark01,puru_prl}
Further experimental efforts for better characterization of the samples and 
further theoretical efforts for detailed study of different types of defects
are needed to reach a quantitative agreement between experiment and 
calculations.
\bibitem{puru_prl}
B. K. Rao and P. Jena, Phys. Rev. Lett. {\bf 89}, 185504 (2002).


\end{thebibliography}
\end{document}